\documentclass[aps,pra,amsmath,twocolumn,amssymb,superscriptaddress]{revtex4-1}

\usepackage{amssymb,amsfonts,amsmath}
\usepackage{color}
\usepackage[apple mac]{inputenc}
\usepackage[english]{babel}
\usepackage{times}
\usepackage{latexsym}
\usepackage{fancyhdr}
\usepackage{verbatim}
\usepackage{graphicx}
\usepackage{epsf}
\usepackage{subfigure}
\usepackage{bm}
\usepackage{epstopdf}

\definecolor{Nathanblue}{rgb}{0.,0.24,0.51}
\newcommand{\blue}{\color{Nathanblue}}

\definecolor{orange}{rgb}{0.96,0.24,0.00}

\usepackage{mathtools}

\def\be{\begin{equation}}
\def\ee{\end{equation}}

\newcommand{\ket}[1]{\ensuremath{\left| #1 \right\rangle}}
\newcommand{\bra}[1]{\ensuremath{\left\langle #1 \right|}}

\begin{document}

\title{{\blue Fractional Chern insulators of few bosons in a box: \\ Hall plateaus from center-of-mass drifts and density profiles}}

\author{C. Repellin}
\email[]{cecile.repellin@lpmmc.cnrs.fr}
\affiliation{Univ. Grenoble-Alpes, CNRS, LPMMC, 38000 Grenoble, France}
\author{J. L\'eonard}
\affiliation{Department of Physics, Harvard University, Cambridge, Massachusetts 02138, USA}
\author{N. Goldman}
\email[]{ngoldman@ulb.ac.be}
\affiliation{CENOLI,
Universit\'e Libre de Bruxelles, CP 231, Campus Plaine, B-1050 Brussels, Belgium}

\begin{abstract}
Realizing strongly-correlated topological phases of ultracold gases is a central goal for ongoing experiments. And while fractional quantum Hall states could soon be implemented in small atomic ensembles, detecting their signatures  in few-particle settings remains a fundamental challenge. In this work, we numerically analyze the center-of-mass Hall drift of a small ensemble of hardcore bosons, initially prepared in the ground state of the Harper-Hofstadter-Hubbard model in a box potential. 
By monitoring the Hall drift upon release, for a wide range of magnetic flux values, we identify an emergent Hall plateau compatible with a fractional Chern insulator state:~the extracted Hall conductivity approaches a fractional value determined by the many-body Chern number, while the width of the plateau agrees with the spectral and topological properties of the prepared ground state. Besides, a direct application of Streda's formula indicates that such Hall plateaus can also be directly obtained from static density-profile measurements. Our calculations suggest that fractional Chern insulators can be detected in cold-atom experiments, using available detection methods.
\end{abstract}

\date{\today}

\maketitle

\section{Introduction} 

Important progress is being made in view of realizing strongly-correlated topological phases of ultracold atoms in optical lattices~\cite{Goldman_topology,cooper2019topological}. On the one hand, experimental efforts have been dedicated to the creation of artificial gauge fields~\cite{Dalibard_review,Goldman_review} and topological bands~\cite{cooper2019topological} for neutral atoms, leading to measurements of topological properties~\cite{atala2013direct,Aidelsburger2015,nakajima2016topological,lohse2016thouless,Wu_2016,sun2018uncover,Flaschner2018,Tarnowski2018,sugawa2018second,lohse2018exploring,meier2018observation,de2019observation,genkina2019imaging,asteria2019measuring,rem2019identifying,chalopin2020exploring,wintersperger2020realization}. On the other hand, theoretical studies have identified realistic schemes for preparing small atomic ensembles in fractional Chern insulators (FCIs)~\cite{cooper2013reaching,yao2013realizing,grusdt2014topological,grusdt2014realization,he-PhysRevB.96.201103,motruk-PhysRevB.96.165107,repellin-PhysRevB.96.161111,hudomal2019bosonic}, which are lattice analogues of fractional quantum Hall (FQH) liquids~\cite{fractional_Chern_insulators, Bergholtz_fci_review}; they also proposed methods to probe their characteristic features~\cite{palmer2006high,kjall2012edge,luo2013edge,Grusdt2016,Taddia2017,dong2018edge,Raciunas2018,umucalilar2018time,repellin2019detecting,Yoshida2020,Cian2020}. This progress should soon lead to the realization of FCIs in small atomic ensembles, with $N\lesssim10$ atoms, and to the possibility of observing their properties. However, identifying clear and accessible topological signatures of FCIs in small interacting atomic systems still constitutes a central challenge. In fact, this question concerns a wide range of quantum-engineered platforms, including strongly-interacting photonic systems~\cite{quantumfluids,ozawa2019topological,RoushanSpectroscopicScience2017,RuichaoDissipativelyNature2019,knuppel2019nonlinear}, where FQH-type states of few photons are currently under intense investigation~\cite{Roushan_2016, AndersonEngineeringPRX2016, ClarkObservationNature2020}.

The canonical signature of FQH states is provided by the Hall conductivity, i.e.~the linear-response coefficient relating an induced transverse current to the applied force. In the FQH effect, the Hall conductivity is quantized to a value $\sigma_\text{H}/\sigma_0\!\in\!\mathbb{Q}$ in the thermodynamic limit~\cite{girvin1999quantum}; $\sigma_0^{-1}\!=\!R_{\text{K}}$ is von Klitzing's constant. The Hall response is also accessible in ultracold atoms; it has been measured in weakly-interacting gases through various probes, including center-of-mass (COM) drifts~\cite{choi2013observation,Jotzu2014,Aidelsburger2015,anderson2019conductivity,genkina2019imaging,wintersperger2020realization} and local  currents~\cite{fletcher2019geometric,chalopin2020exploring}, and more indirectly, through collective-mode excitations~\cite{leblanc2012observation} and circular dichroism~\cite{asteria2019measuring}. Whether the Hall response can be extracted and used as a topological marker in few-body interacting systems remain important questions to be addressed.

\begin{figure}[h!]
\includegraphics[width = \linewidth]{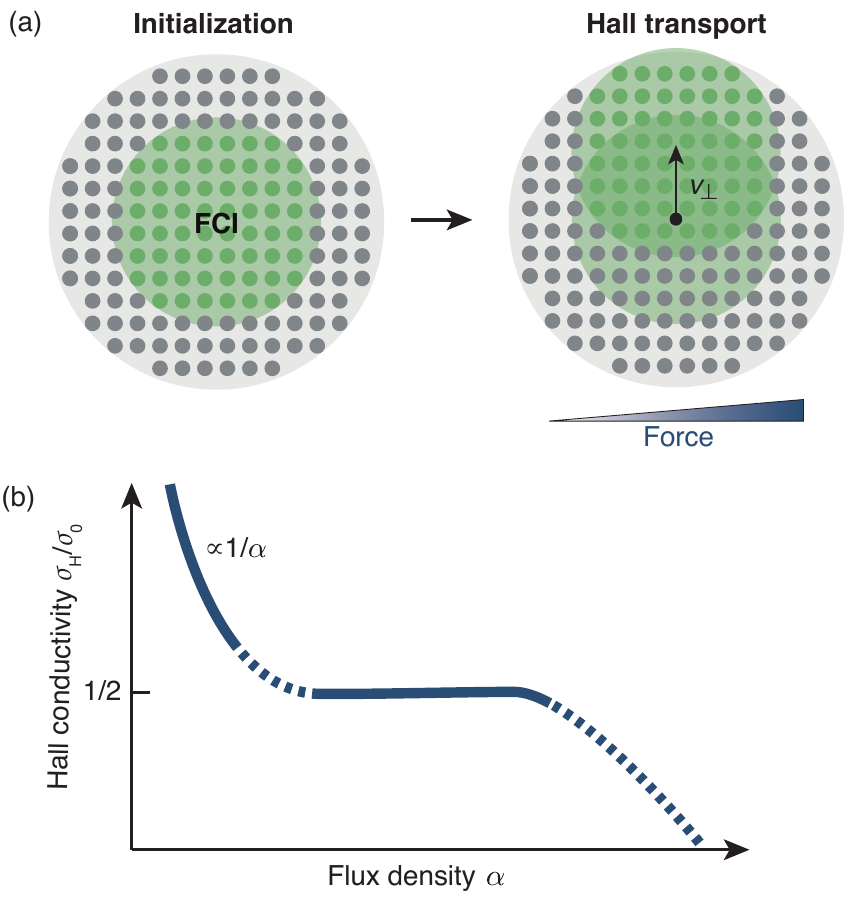}
\caption{(a) Hall drift protocol:~a prepared fractional Chern insulator (FCI) state is released into a larger lattice and a uniform force is applied. The Hall conductivity is extracted from the COM drift transverse to the force [Eq.~\ref{eq:extract}]. (b) Sketch of the Hall conductivity $\sigma_\text{H}(\alpha)$ as a function of the flux density in the Harper-Hofstadter-Hubbard (HHH) model. In the continuum limit $\alpha\!\ll\!1$, the system follows the classical prediction $\sigma_\text{H}\!\sim\!1/\alpha$. In the vicinity of the filling factor $\nu \!=\!\rho_{\text{bulk}}/\alpha\!=\!1/2$, where $\rho_{\text{bulk}}$ is the bulk particle density, an FCI is formed and $\sigma_\text{H}(\alpha)$ depicts a Hall plateau.
}
\label{fig1}
\end{figure}

In this work, we numerically analyze the Hall drift of a small ensemble of strongly-interacting (hardcore) bosons, initially prepared in the ground state of the Harper-Hofstadter-Hubbard model~\cite{hofstadter1976energy,sorensen-PhysRevLett.94.086803}. Building on Refs.~\cite{dauphin2013,Aidelsburger2015}, we monitor the COM of the prepared state upon releasing it into a larger lattice while applying a weak static force [Fig.~\ref{fig1}(a)]. This Hall drift measurement, which provides an estimation of the Hall conductivity in the prepared state, is performed in a wide range of magnetic flux values [Fig.~\ref{fig1}(b)].
From this, we identify an emergent but robust Hall plateau, whose value $\sigma_{\text{H}}/\sigma_0\!\approx0.5$ approaches the many-body Chern number~\cite{NiuThouless,tao1986impurity,hafezi-PhysRevA.76.023613}, a topological marker of the FCI phase. Moreover, the width of the Hall plateau perfectly coincides with the flux window where an FCI state is formed in the initially confined geometry, as we demonstrate based on a static analysis of the ground-state's entanglement spectrum. Our results indicate that Hall signatures of FCI states composed of few bosons ($N\!\ge\!3$) can be identified under realistic experimental conditions. 
We also compare this approach with a direct application of Streda's formula~\cite{Widom,Streda,Streda2,giuliani2005quantum}, which indicates that static density-profile measurements lead to a robust Hall plateau for $N\!\ge\!10$ bosons.

This article is organized as follows: We first analyze the ground-state properties of our model in Section~\ref{section:GS}, setting the focus on the subtle identification of FCI states in confined systems with edges. We then study the center-of-mass Hall drift of this setting, upon release into a larger lattice, and comment on the emergence of quantized Hall plateaus in Section~\ref{section:drift}. We then explore the applicability of the Streda-formula approach in Section~\ref{section:streda}, before concluding with a discussion on the experimental realization of our Hall drift protocol in Section~\ref{section:conclusions}.

\section{Fractional Chern insulator in a box: Ground-state properties}\label{section:GS} 

The central scope of this work concerns the emergence of quantized Hall plateaus in the COM dynamics of strongly-interacting bosons moving on a 2D square lattice in the presence of a uniform magnetic flux.
The corresponding Harper-Hofstadter-Hubbard (HHH) Hamiltonian reads~\cite{hofstadter1976energy,sorensen-PhysRevLett.94.086803} 
\begin{align}
\hat{H}_0=&-J\left(\sum_{m, n} \hat{a}^\dagger_{m, n+1} \hat{a}_{m,n} + e^{i 2 \pi \alpha n}\hat{a}^{\dagger}_{m+1, n}\hat{a}_{m,n} + \text{h.c.} \right) \notag \\
&+(U/2) \sum_m \hat{a}^\dagger_{m, n}\hat{a}_{m,n} (\hat{a}^\dagger_{m, n}\hat{a}_{m,n}-1),
\label{eq: time independent H}
\end{align}
where $\hat{a}^\dagger_{m, n}$ creates a boson at lattice site $(m,n)$, $J$ denotes the tunneling amplitude, $U$ is the on-site (Hubbard) interaction strength~\cite{bloch2008many}, and where the Peierls phase factors~\cite{hofstadter1976energy} account for the presence of a flux $\Phi\!=\!2 \pi \alpha$ per plaquette. This model has been experimentally implemented for $N\!=\!2$ strongly-interacting bosonic atoms~\cite{Tai2017}, in a box-type potential~\cite{BauntBosePRL2013, ChomazEmergenceNatComm2015, Tai2017, ChiuQuantumPRL2018, SaintJalmDynamicalPRX2019}. It is the aim of this Section to shed some light on the ground-state properties of this realistic system for $N>2$ bosons.

Numerical simulations using periodic boundary conditions have established that the HHH model hosts a bosonic FCI akin to the Laughlin state~\cite{girvin1999quantum}, for strongly repulsive interactions and filling factor $\nu \!=\!\rho/\alpha\!=\!1/2$, where $\rho$ denotes the particle density; see Refs.~\cite{sorensen-PhysRevLett.94.086803, palmer2006high, hafezi-PhysRevA.76.023613, repellin-PhysRevB.96.161111, he-PhysRevB.96.201103, motruk-PhysRevB.96.165107}. For hardcore bosons, these calculations reveal a stable FCI ground state for $\alpha\!\leq\!0.3$; the bulk gap is maximal around $\alpha\!\approx\!0.2-0.25$~\cite{sorensen-PhysRevLett.94.086803, hafezi-PhysRevA.76.023613, repellin-PhysRevB.96.161111}, and vanishes in the limit $\alpha\!\ll \!1$. This FCI phase is characterized by a fractional many-body Chern number, $\nu_{\text{Ch}}^{\text{MB}}\!=\!1/2$, a topological invariant associated with the ground-state of the many-body system~\cite{NiuThouless,tao1986impurity,hafezi-PhysRevA.76.023613}. In the thermodynamic limit, the Hall conductivity of an incompressible phase approaches the quantized value $\sigma_{\text{H}}/\sigma_0\!=\!\nu_{\text{Ch}}^{\text{MB}}\!\in\!\mathbb{Q}$; see Ref.~\cite{repellin2019detecting} for a numerical analysis of finite-size effects. In the relevant case of the $\nu\!=\!1/2$ FCI phase, a Hall plateau is thus expected at $\sigma_{\text{H}}/\sigma_0\!=\!1/2$; see Fig.~\ref{fig1}(b).

In the present work, we consider $N$ hardcore bosons initially confined in a circular box containing $N_s$ lattice sites [Fig.~\ref{fig1}(a)]; such box potentials are indeed available in experiments~\cite{BauntBosePRL2013, ChomazEmergenceNatComm2015, Tai2017, ChiuQuantumPRL2018, SaintJalmDynamicalPRX2019}. In this geometry with edges, we still expect to find the $\nu\!=\!1/2$ FCI phase in the regime $\rho_{\text{bulk}}/\alpha\!\approx\!1/2$, where $\rho_{\text{bulk}}$ denotes the bulk particle density. However, we note that the bulk density $\rho_{\text{bulk}}$, which differs from the total density $\rho\!=\!N/N_s$, has a non-trivial dependence on the flux; see Section~\ref{section:streda}. Hence, we first analyze the ground-state properties of this few-body system in view of determining the values of $\rho$ and $\alpha$ that realize an FCI state. 

We first gain some intuition from the physics of the FQH effect in the (continuum) disk geometry~\footnote{We refer to the infinite plane geometry, where the FQH droplet is confined through total angular momentum conservation, not by a confinement potential.}. For a flux density $\alpha\!\leq\!0.3$, the lowest Bloch band of the single-particle Hofstadter spectrum contains a set of roughly $N_0 (\alpha)$ nearly degenerate states, which are connected to the next band by dispersive edge states; see Appendix~\ref{app:spectrum}. These $N_0 (\alpha)$ states are analogous~\cite{kjall2012edge,luo2013edge} to the orbitals of the lowest Landau level (LLL) in the disk geometry; there, the $\nu\!=\!1/2$ Laughlin state with $N$ bosons occupies $2N\!-\!1$ LLL orbitals. Likewise, we may expect a $\nu\!=\!1/2$ FCI state when $N_0(\alpha)\!\simeq\!2N\!-\!1$. 

\begin{figure}[h!]
\includegraphics[width = \linewidth]{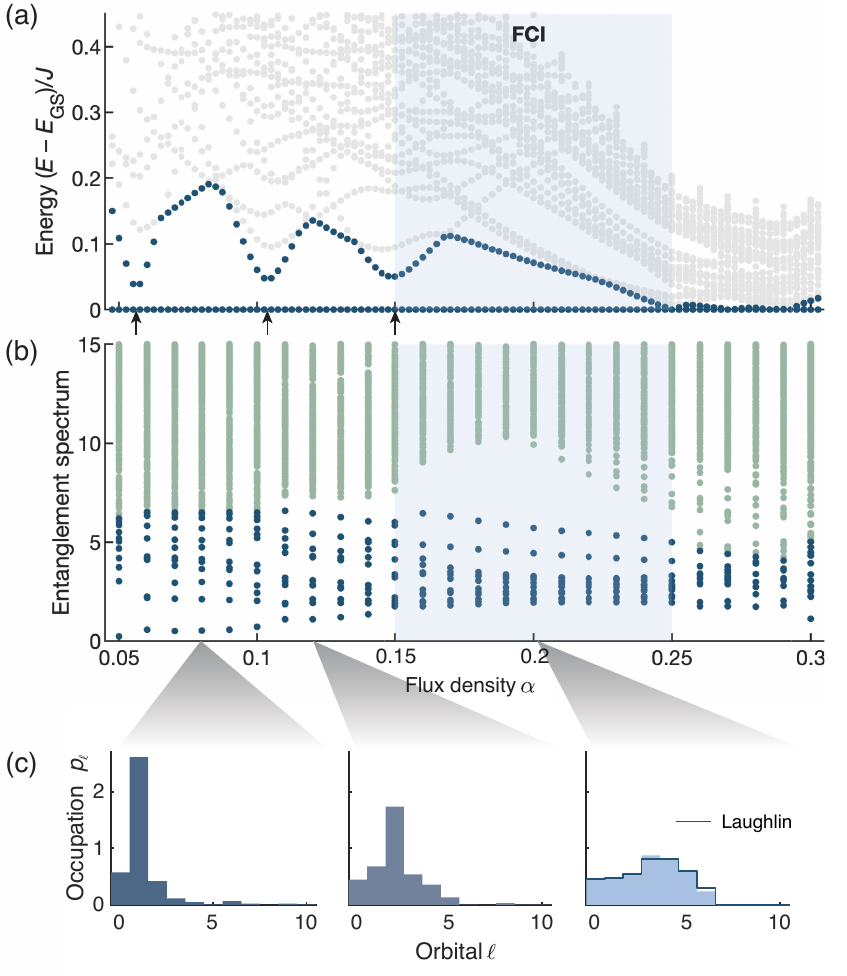}
\caption{Spectral and topological ground-state properties. We present the ground-state properties for $N\!=\!4$ hardcore bosons in the HHH model, in a circular box of $N_s\!=\!60$ sites. The FCI stability region (shaded) is indicated according to three markers:~(a) Low-energy many-body spectrum (lowest~$10$ energies per discrete rotation-symmetry sector) relative to the ground-state energy $E_{\text{GS}}$. Analyzing absolute energies (not shown) reveals that the three local minima of the gap correspond to level crossings, which are avoided due to finite-size effects (arrows).~(b) Particle entanglement spectrum (PES) for a bipartition with~$2$ particles in each subsystem. In the shaded region, the first~$15$ levels (blue) are well separated from the other levels, revealing that the ground state is topologically equivalent to the Laughlin state.~(c) Occupation of the single-particle orbitals in the ground state (histograms) and in the exact Laughlin state on the disk (line). The orbitals are sorted in increasing energy and angular momentum, respectively.}
\label{fig2}
\end{figure}

We use exact diagonalization to verify the existence of the FCI ground state in our model, and specify its phase boundaries based on (i)~the low-energy spectrum; (ii)~entanglement spectroscopy; (iii)~the occupation of single-particle orbitals.
For concreteness, we analyse $N\!=\!4$ hardcore bosons in $N_s\!=\!60$ sites:

(i) Figure~\ref{fig2}(a) shows the low-energy spectrum of this few-body system; there are three avoided crossings between the ground state and the first excited state  within the range $0\!\leq\!\alpha\!\leq\!0.3$, which we  interpret as the finite-size signatures of three phase transitions.
We focus on the regime $0.15\!<\!\alpha\!<\!0.25$, where the expected FCI bulk gap is the largest~\cite{sorensen-PhysRevLett.94.086803, repellin-PhysRevB.96.161111}, and no phase transition is observed. In this regime, the approximate degeneracy of the lowest band is $N_0(\alpha)\!\simeq\!7$ [see Appendix~\ref{app:spectrum}], which is compatible with an FCI ground state candidate for $N\!=\!4$. We note that the nature of the phases at $\alpha\!<\!0.15$ is likely to be non-universal due to finite-size effects~\footnote{The $\nu\!=\!2/3$ Jain fraction is a possible candidate among these phases~\cite{moller-PhysRevLett.103.105303, kjall2012edge}, but we did not identify any clear topological signature of this phase in our setting.}. 

(ii) In finite geometries with edges, the topological nature of FCIs can be revealed through the degeneracies of their edge spectrum~\cite{kjall2012edge,luo2013edge}; however, this spectral signature requires a smooth confining potential~\footnote{In the continuum, a hard confinement was shown to suppress the typical Luttinger liquid dispersion relation of edge modes~\cite{fern2017hardconfinement, macaluso2017hardwall}}. Instead, we probe the bulk quasihole excitations of the ground state; their degeneracy is a topological fingerprint of the FCI phase, and it can be extracted from the ground state $\ket{\Psi_{\mathrm{GS}}}$ using the particle entanglement spectrum~\cite{sterdyniak-PhysRevLett.106.100405} (PES). The PES is the spectrum of the reduced density matrix obtained by tracing  $\ket{\Psi_{\mathrm{GS}}} \bra{\Psi_{\mathrm{GS}}}$ over a bipartition containing $N_A$ particles, while keeping the geometry of the system intact. The degeneracy of Laughlin quasiholes is determined through a generalized exclusion principle~\cite{haldane1991fractional}; for $2$ bosons in $7$ orbitals, it is $15$-fold. And indeed, for $0.15\!<\!\alpha\!<\!0.25$ and $N_A\!=\!2$, the PES in Fig.~\ref{fig2}(b) reveals a clear gap above the 15th state, which confirms the FCI nature of $\ket{\Psi_{\mathrm{GS}}}$ in this parameter range.

(iii) To further characterize the ground state, we calculated the occupation of each single-particle orbital. For the Laughlin state on the disk, there is a uniform occupation of all $2N-1$ orbitals in the thermodynamic limit, with moderate deviations from this distribution for small systems. We find a similar distribution in the regime $0.15\!\lesssim\! \alpha\!\lesssim\!0.25$ for the same number of particles [Fig.~\ref{fig2}(c)].

Overall, these probes consistently reveal that the ground state is in the $\nu\!=\!1/2$ FCI phase within the range $0.15\!\lesssim\!\alpha\!\lesssim\!0.25$. 
We have verified that these results are qualitatively robust with respect to changes in the particle number ($N\!=\!3, 4$) and number of sites $N_s$. Quantitatively, we point out that a change in the density $\rho$ leads to a shift of the FCI phase along the flux axis; see Appendix~\ref{section:sizes}.

\section{The center-of-mass Hall drift} \label{section:drift}

We extract the Hall conductivity $\sigma_{\text{H}}$ from the COM drift of an initially prepared state upon applying a weak external force. This COM probe~\cite{dauphin2013,Aidelsburger2015,price2016measurement} is particularly suitable when considering a small ensemble of particles, for which local currents substantially fluctuate. In order to limit boundary effects, we release the initially prepared state into a larger lattice~\cite{dauphin2013,Tran2017} before monitoring the COM drift; see Fig.~\ref{fig1}(a). This protocol is implementable in cold-atom experiments, where tunneling can be dynamically tuned and COM motion measured~\cite{choi2013observation,Jotzu2014,Aidelsburger2015,anderson2019conductivity,genkina2019imaging,wintersperger2020realization}. In our simulations, the time scale associated with the progressive release of the inner system, as well as the duration of the Hall-drift measurement, are adjusted to avoid boundary effects due to the finite simulation box.

We extract the Hall velocity $v_{\perp}\!=\!x_{\perp}(t)/t$ from the COM Hall drift $x_{\perp}(t)$ transverse to the applied force $F$, upon reaching a stationary regime within linear response. This Hall velocity is related to the transverse current density through $v_{\perp}\!=\!j_{\perp}/\rho_{\text{bulk}}$, where $\rho_{\text{bulk}}$ denotes the bulk particle density. From the transport equation, $j_{\perp}\!=\!\sigma_{\text{H}} F$, one extracts the Hall conductivity through 
\begin{equation}
\sigma_{\text{H}}/\sigma_0\!=\!(2 \pi \rho_{\text{bulk}}/F)v_{\perp},\label{eq:extract}
\end{equation}
where $\sigma_0\!=\!1/2\pi$ is the conductivity quantum, and we set $\hbar\!=\!1$ except otherwise stated. In the FQH effect~\cite{giuliani2005quantum,mudry2014lecture}, incompressible states exhibit quantized Hall plateaus at fractional values $\sigma_{\text{H}}/\sigma_0\!=\!\nu_{\text{Ch}}^{\text{MB}}\!\in\!\mathbb{Q}$, where $\nu_{\text{Ch}}^{\text{MB}}$ denotes the aforementioned many-body Chern number~\cite{NiuThouless,tao1986impurity}. The main scope of this Section is to analyze the possibility of observing such Hall plateaus, through the Hall drift of few-boson FCI states.

\subsection{Benchmark using non-interacting fermions} 

We first benchmark the Hall-drift measurement by considering non-interacting fermions in the Harper-Hofstadter model, at quarter filling $\rho\!=\!1/4$. In this setting, (integer) Chern insulators are expected around flux densities $\alpha\!=\!1/(4 \mathfrak{n})$, with $\mathfrak{n}\!\in\mathbb{Z}$, where they exhibit quantized Hall plateaus $\sigma_\text{H}/\sigma_0\!=\!\mathfrak{n}$ in the thermodynamic limit. While this result can be directly deduced from a Diophantine equation~\cite{thouless1982quantized,kohmoto1989zero}, the actual size of these plateaus follows a rather complicated law established by the underlying single-particle (Hofstadter-butterfly) spectrum~\cite{hofstadter1976energy}. Furthermore, such Hall plateaus can be altered by finite-size effects. This first numerical study aims to shed some light on these properties.

\begin{figure}[h!]
\includegraphics[width = \linewidth]{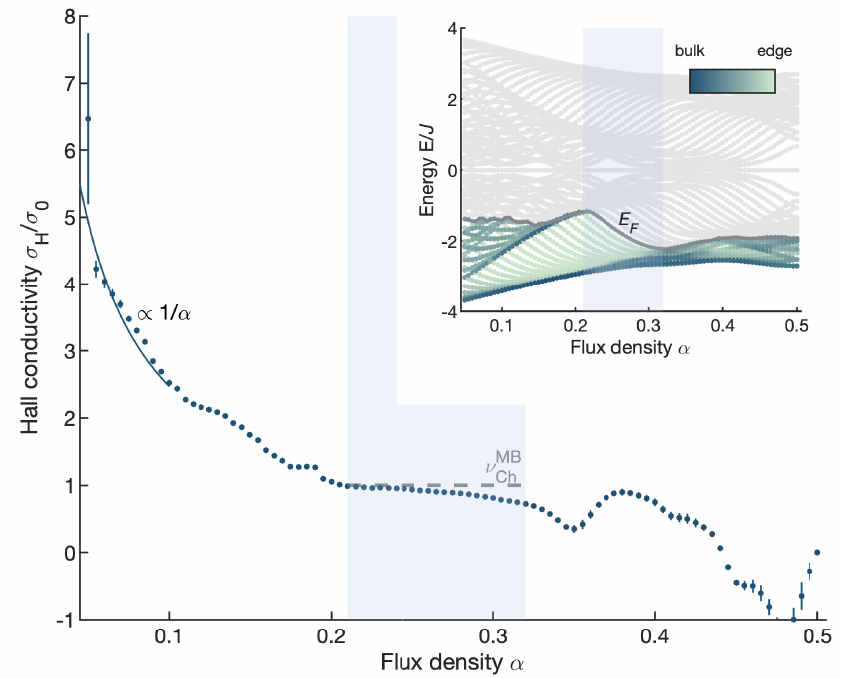}
\caption{Benchmark using non-interacting fermions.~Hall conductivity extracted from the Hall drift of $N\!=\!20$ fermions, initially prepared in a small circular box of $N_s\!=\!81$ sites (quarter filling); the release into the larger system ($1005$ sites) and the ramping up of the force ($F\!=\!0.2 J/d$) are performed over a duration $\tau_{\text{ramp}}\!=\!15J^{-1}$, and the Hall drift measurement over $\tau_{\text{hold}}\!=\!13J^{-1}$; the bulk density $\rho_{\text{bulk}}(\alpha)$ in Eq.~\eqref{eq:extract} is evaluated in a central region of 20 sites; 
fit error bars reflect the 95$\%$ confidence interval when extracting the Hall velocity $v_{\perp}$. The gray dashed line indicates the quantized value expected in the thermodynamic limit, $\sigma_\text{H}/\sigma_0\!=\!1$, as dictated by the many-body Chern number of the corresponding insulating state. (Inset) Single-particle spectrum $E(\alpha)$, for the same box ($81$ sites); the Fermi level ($E_F$; grey curve) corresponds to the same number of fermions $N\!=\!20$; filled states are colored in green (bulk states are dark and edge states are light), while empty states are grey; $E_F$ is located in the main bulk gap (Chern insulator) within the blue shaded region.}
\label{fig3}
\end{figure}

We determine the Hall drift by calculating the time-evolution of $N\!=\!20$ non-interacting fermions, initially prepared in the ground-state within a circular box of $N_s\!=\!81$ sites ($\rho\!\approx\!1/4$), which are then released into a larger lattice ($1005$ sites) and subjected to a weak force $F\!=\!0.2 J/d$, where $d$ is the lattice spacing; see Appendix~\ref{methods}. The Hall conductivity extracted from the stationary (linear-response) Hall drift [Eq.~\eqref{eq:extract}] is depicted as a function of the flux density in Fig.~\ref{fig3}. In the low-flux regime ($\alpha\!\lesssim\!0.1$), lattice effects are negligible and the Hall drift follows the classical prediction $\sigma_{\text{H}}\!\sim\! 1 / \alpha$. At $\alpha\!=\!0.25$, the Fermi energy lies within a large bulk gap (see inset of Fig.~\ref{fig3}), which yields an approximately quantized value $\sigma_\text{H}/\sigma_0\!\approx\!0.9$, close to the many-body Chern number $\nu_{\text{Ch}}^{\text{MB}}\!=\!1$ of the corresponding insulating state~\cite{NiuThouless}. In order to explain the deviation from the exact quantized value expected in the thermodynamic limit, we have evaluated the local real-space Chern number~\cite{bianco2011mapping,tran2015topological}, as averaged over 29 bulk sites in the prepared ground state; we have verified that this local real-space Chern number indeed matches the value of the extracted Hall conductivity $\sigma_\text{H}/\sigma_0\!\approx\!0.9$; see Appendix~\ref{app:local} regarding the convergence of the local real-space Chern number as a function of the system size.

The Hall response remains approximately constant for a wide range of flux, $\alpha\!\in\![0.21 , 0.32]$, in agreement with the Fermi level's location within the main bulk gap of the butterfly spectrum [Fig.~\ref{fig3}]; we verified that the flatness of this emergent Hall plateau, as well as the approached quantized value, improve as the system size $N_s$ increases. Besides, we also observe the emergence of additional plateaus in Fig.~\ref{fig3}, as the Fermi level visits other bulk gaps in the spectrum. These results illustrate how the Hofstadter-butterfly spectrum dictates the size of emergent Hall plateaus in realistic finite-size (non-interacting) settings.

\subsection{Hall drift of interacting bosons} 
\begin{figure}[h!]
\includegraphics[width = \linewidth]{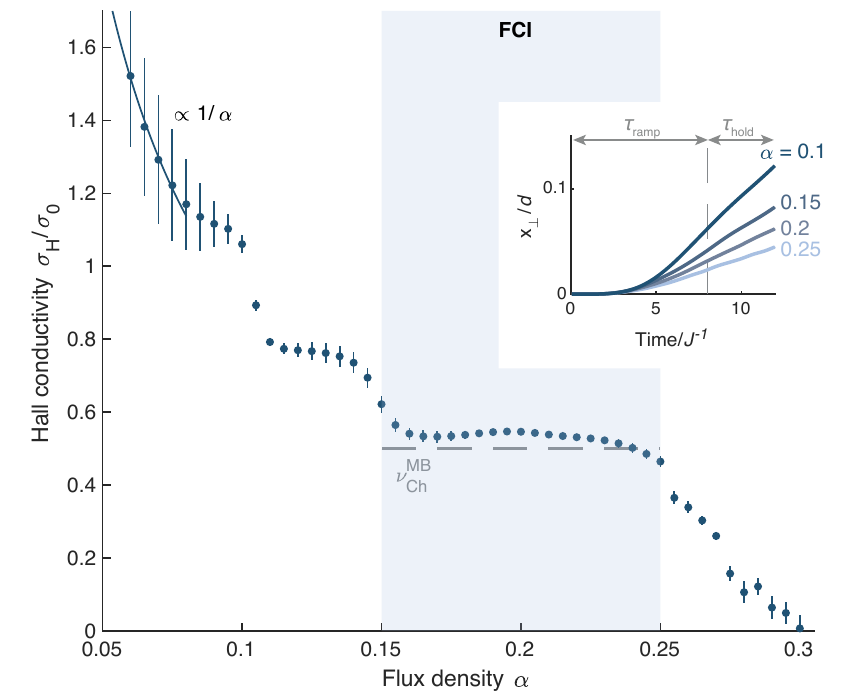}
\caption{Hall drift of hardcore bosons in the HHH model. Initially, $N\!=\!4$ bosons are prepared in the ground state within a circular box of $N_s\!=\!60$ sites. The bosons are then progressively released into a larger box containing $124$ sites and subjected to a uniform force $F\!=\!0.01 J/d$. At the end of this ramp, the COM motion is stationary for a few tunneling times (inset), permitting the extraction of the Hall velocity $v_{\perp}$; the bulk density $\rho_{\text{bulk}}(\alpha)$ is evaluated in a central region of 16 sites; fit error bars reflect the 95$\%$ confidence interval when extracting the Hall velocity $v_{\perp}$. The FCI phase (shaded region from Fig.~\ref{fig2}) coincides with the emergent Hall plateau. The gray dashed line indicates the quantized value expected in the thermodynamic limit, $\sigma_\text{H}/\sigma_0\!=\!1/2$, as dictated by the many-body Chern number.}
\label{fig4}
\end{figure}

A system of $N\!=\!4$ hardcore bosons is initially prepared in the ground state of the HHH model, using a circular box  of $N_s\!=\!60$ sites. At $t\!=\!0$, this bosonic cloud is slowly released into a larger circle ($124$ sites), while the force is ramped up to the value $F\!=\!0.01 J/d$; see Appendix~\ref{methods} for details on the ramps used to reach a stationary regime. 
Our numerics show that a stationary COM motion takes place after the ramp, over a few tunneling times [inset of Fig.~\ref{fig4}], from which we extract the (constant) Hall velocity $v_{\perp}$. For longer times, the COM motion is affected by the edge of the finite simulation box, which sets the end of the stationary regime; we point out that this numerical constraint is not an experimental one, since the prepared state can be released into a much larger lattice in realistic setups. 

We extract the Hall conductivity $\sigma_{\text{H}}$, from the stationary Hall velocity $v_{\perp}$ and bulk density $\rho_{\text{bulk}}$ [Eq.~\eqref{eq:extract}], for a large range of flux values; see Fig.~\ref{fig4}. First, we find that the classical behavior $\sigma_{\text{H}}\!\sim\! 1 / \alpha$ is recovered~\cite{lindner2009vortex} in the low-flux (continuum) limit. The correlated behavior of our interacting system then appears for $\alpha\!\geq\!0.1$. Most strikingly, $\sigma_{\text{H}}(\alpha)$ depicts an emergent Hall plateau, whose width matches the flux window associated with the $\nu\!=\!1/2$ FCI phase [Fig.~\ref{fig2}]. In this region, the extracted Hall conductivity approaches the quantized value $\sigma_{\text{H}}/\sigma_0\!\approx\!0.5$, which is expected in the thermodynamic limit. Calculations performed on larger systems (up to $N\!=\!10$ bosons in $N_s\!=\!120$ sites) indeed demonstrate convergence to a quantized Hall plateau; see Appendix~\ref{section:sizes}. Interestingly, a second plateau appears in the range $\alpha\!\in\![0.1, 0.15]$, which may signal the onset of the $\nu\!=\!2/3$ FCI phase~\cite{moller-PhysRevLett.103.105303, kjall2012edge}.

In the prospect of experimental implementation, we sought for minimal configurations that exhibit Hall-plateau signatures of the FCI state, by simulating our Hall drift protocol for various system sizes $N_s$, considering $N\!=\!3, 4$ bosons; see Appendix~\ref{section:sizes}. Our results show that emergent Hall plateaus, compatible with the $\nu\!=\!1/2$ FCI state, can be detected by measuring the COM Hall drift of systems as small as $3$ bosons in $N_s\!=\!40$ sites and $4$ bosons in $N_s\!=\!49$ sites. 

\subsection{Stability of the Hall plateau and bulk density evaluation} \label{subsec: stability}

Extracting the Hall conductivity $\sigma_{\text{H}}$ from the center-of-mass (COM) Hall drift requires the evaluation of the bulk density $\rho_{\text{bulk}}$, according to Eq.~\eqref{eq:extract}. For large enough systems, one can evaluate $\rho_{\text{bulk}}$ by averaging  the density over a central circular region, whose radius is large compared to the lattice spacing but small compared to the radius of the atomic cloud; see Fig.~\ref{figStreda} and Section~\ref{section:streda}. In few-boson systems, however, such scale separation may not exist. For instance, in the configuration studied in Fig.~\ref{fig4}, i.e.~$N\!=\!4$ bosons initially confined in $N_s\!=\!60$ sites, one notices substantial spatial fluctuations of the particle density. 

\begin{figure}[h!]
\includegraphics[width = \linewidth]{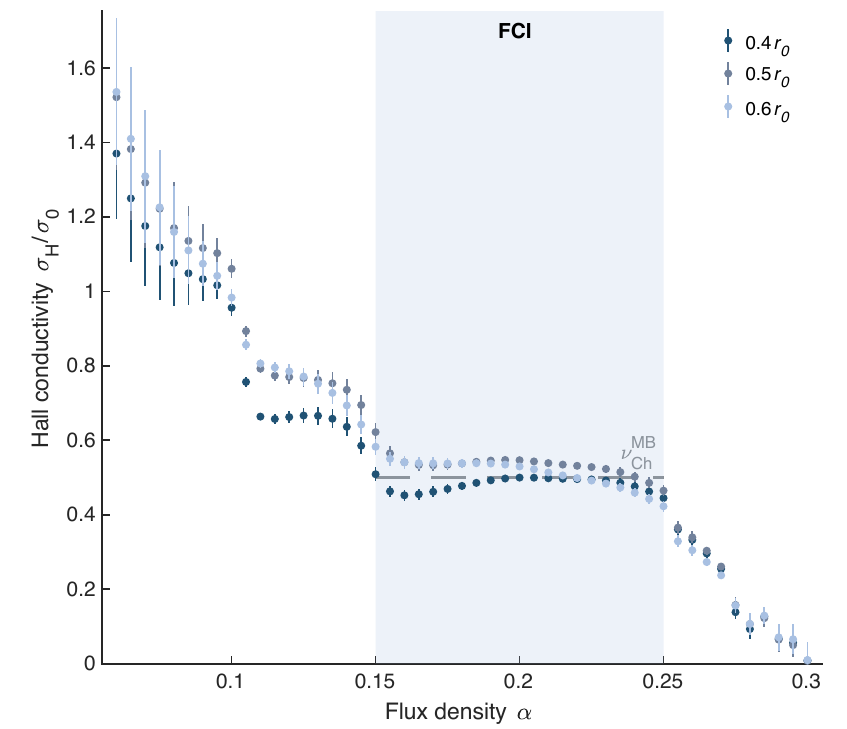}
\caption{Impact of the bulk density evaluation on the determination of $\sigma_{\text{H}}$ for the few-boson system in Fig.~4. Here, we determined $\rho_{\text{bulk}}$ as the average particle density within a central circular region of radius $r_{\text{bulk}}$, with $r_{\text{bulk}}\!=\!0.4 r_0$, $r_{\text{bulk}}\!=\!0.5 r_0$ and $r_{\text{bulk}}\!=\!0.6 r_0$, where $r_0$ is the radius of the initial circular box ($N_s\!=\!60$); these three radii correspond respectively to $12$, $16$ and $24$ sites. In this small system, the radius choice affects the value of $\sigma_{\text{H}}$, as extracted from the COM Hall drift [Eq.~\eqref{eq:extract}]; however, the existence of the Hall plateau, as well as its position and range along the flux axis, are robust.
}
\label{figsuppImpactBulkDensity}
\end{figure}

Here, we analyze the impact of such density fluctuations on the extracted Hall conductivity shown in Fig.~\ref{fig4}. To do so, we evaluate $\rho_{\text{bulk}}$ as the average particle density within a small and central circular region of radius $r_{\text{bulk}}$, considering values in the range $r_{\text{bulk}}\!\in\![0.4r_0 , 0.6r_0 ]$, where $r_0$ is the radius of the (initial) circular box containing $N_s$ sites. We illustrate the impact of this radius choice in Fig.~\ref{figsuppImpactBulkDensity}, which shows $\sigma_{\text{H}}$ as extracted from the COM Hall drift [Eq.~\eqref{eq:extract}] for three different choices. Interestingly, the Hall plateau displayed in Fig.~\ref{fig4} is shown to be very robust:~neither its existence, nor its position and range on the flux axis, depend on the radius $r_{\text{bulk}}$ used to evaluate the bulk density. While the exact value of $\sigma_{\text{H}}/\sigma_0$ depends on $r_{\text{bulk}}$, one notices that $\sigma_{\text{H}}/\sigma_0\!\approx\!0.5$ on this plateau for all choices.\\

\section{Extracting the Hall plateau from the particle density:~Streda's formula}\label{section:streda}

In this Section, we consider an alternative approach based on density-profile measurements. For an incompressible phase, the variation of the bulk density $\rho_{\text{bulk}}$ with respect to the flux density $\alpha$ is directly related to the quantized Hall conductivity. This relation, which is known as Streda's formula~\cite{Widom,Streda,Streda2,giuliani2005quantum,price2016measurement,Oktel,Fukui_Streda}, reads in our units
\begin{equation}
\sigma_{\text{H}}/\sigma_0\!=\!\frac{\partial \rho_{\text{bulk}}}{\partial \alpha}.\label{streda} 
\end{equation}
Applying this approach to the $\nu\!=\!1/2$ FCI phase, one thus expects the existence of a plateau $\partial \rho_{\text{bulk}}/\partial \alpha \!\approx\!0.5$ within the corresponding flux range. 

\begin{figure}[h!]
\includegraphics[width = \linewidth]{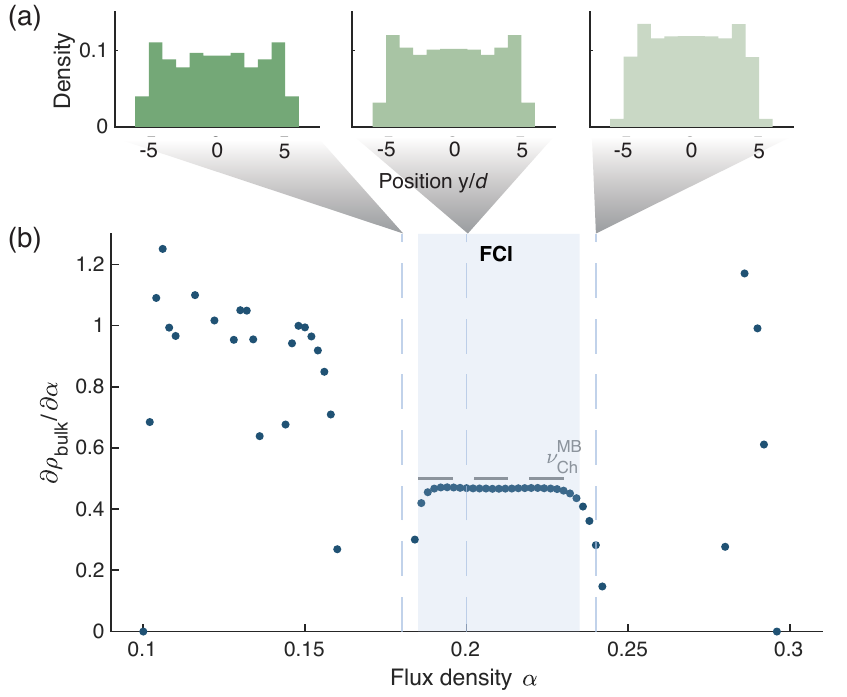}
\caption{Streda analysis of the HHH ground-state, for $N\!=\!10$ hardcore bosons in a circular box of $N_s=120$ sites, using DMRG: (a) Particle density profiles; (b) Derivative of the bulk particle density $\rho_{\text{bulk}}$ with respect to the flux density; in the FCI phase, it satisfies Streda's formula in Eq.~\eqref{streda}. The bulk density $\rho_{\text{bulk}}$ was evaluated as the average density in a circle comprising $16$ sites. The gray dashed line indicates the quantized value expected in the thermodynamic limit, $\sigma_{\text{H}}/\sigma_0\!=\!\partial \rho_{\text{bulk}}/\partial \alpha=1/2$, as dictated by the many-body Chern number.
}
\label{figStreda}
\end{figure}

We have validated this prediction by analyzing the density profiles of $N\!=\!10$ hardcore bosons confined in $N_s\!=\!120$ sites, for various values of the flux $\alpha$. We display three representative profiles in Fig.~\ref{figStreda} (a), which illustrate the existence of a density plateau in the bulk. One verifies that the bulk density $\rho_{\text{bulk}}$, as extracted from the density plateau, increases linearly as a function of the flux $\alpha$ within a well-defined flux window. This result is represented in Fig.~\ref{figStreda} (b), which depicts the derivative of the extracted bulk density [Eq.~\eqref{streda}]. The clear plateau at $\sigma_{\text{H}}/\sigma_0\!=\!\frac{\partial \rho_{\text{bulk}}}{\partial \alpha}\approx1/2$ offers a striking signature of the FCI phase within this flux window. One verifies that this ``Streda-Hall" plateau matches the plateau obtained from the Hall drift protocol using the same system parameters; see Fig.~\ref{figsuppDMRG} in Appendix~\ref{section:sizes}.

For smaller systems, the irregular density profile complicates the extraction of $\rho_{\text{bulk}}$, as already explained in Section~\ref{subsec: stability}. Nevertheless, evaluating $\rho_{\text{bulk}}$ in a small circular region of radius $r_{\text{bulk}}$, we still observe emergent plateaus of $\partial \rho_{\text{bulk}}/\partial \alpha$ within the FCI phase of smaller systems ($N\!\sim\!4$). In those configurations, the value on emergent plateaus displays a strong dependence on $r_{\text{bulk}}$, which complicates a quantitative application of Streda's formula for such small systems.

\section{Experimental considerations}\label{section:conclusions}

For a possible implementation of the proposed Hall-drift protocol, we consider the following experimental scheme. First, the FCI is prepared through adiabatic quantum state engineering, starting from a topologically trivial state and inducing a topological phase transition to an FCI by slowly tuning the Hamiltonian parameters~\cite{grusdt2014topological,grusdt2014realization,he-PhysRevB.96.201103,motruk-PhysRevB.96.165107,repellin-PhysRevB.96.161111,hudomal2019bosonic}. The adiabaticity of this preparation relies on the small number of particles and size of the system, which prevents the many-body gap from vanishing at the transition point. In our scheme, the FCI would be embedded in a lattice with more sites, which are initially uncoupled either by switching off the tunneling to those sites (as assumed in our numerical calculations), or by increasing their energy with a repulsive potential. The drift protocol is initiated by restoring the coupling to outer sites and simultaneously ramping up a force induced by, for instance, an optical or magnetic potential gradient~\cite{Jotzu2014,Aidelsburger2015}. Finally, the Hall drift is detected by measuring the COM position after variable drift times. Detecting COM displacements smaller than one lattice site, as depicted in Fig.~\ref{fig4}, is within the current capabilities of cold-atom experiments~\cite{Tai2017}. Yet, we expect that even stronger signatures are possible because experiments allow access to total system sizes and drift times beyond the reach of exact numerics. In addition, the ability to choose finite interactions and to tune them dynamically opens up the possibility for advanced transport studies. The simplicity of this realistic experimental scheme paves the way to the exploration of quantum transport in strongly-correlated ultracold topological matter.\\

During completion of our manuscript, we became aware of
a recent work ~\cite{Motruk_prep}, which also analyses the Hall response of an FCI in the HHH model. \\

\paragraph*{Acknowledgments} 
We thank M.~Greiner for insightful discussions, and I.~Carusotto and M.~Hafezi for comments.
We acknowledge J. Motruk and I. Na for sharing their manuscript~\cite{Motruk_prep} before submission, and for their comments on our work. DMRG calculations were performed using the TeNPy Library (version 0.4.0)~\cite{package}. C.R. is supported by the Marie Sklodowska-Curie program under EC Grant agreement 751859. J. L. acknowledges support from the Swiss National Science Foundation. N.G. is supported by the FRS-FNRS (Belgium) and the ERC Starting Grant TopoCold.

\bibliography{FCICOM}
\bibliographystyle{ieeetr}

\clearpage
\newpage

\appendix

\twocolumngrid

\section{Other system sizes}\label{section:sizes}
In the main text, we have shown numerical data for $N\!=\!4$ hardcore bosons prepared in a circular box of $N_s\!=\!60$ sites. We have obtained consistent results for other particle numbers $N$ and lattice sizes $N_s$, (corresponding to other densities $\rho$), which we present in this Appendix.

\subsection{Hall drift in larger systems: DMRG results}
\begin{figure}[h!]
\includegraphics[width = \linewidth]{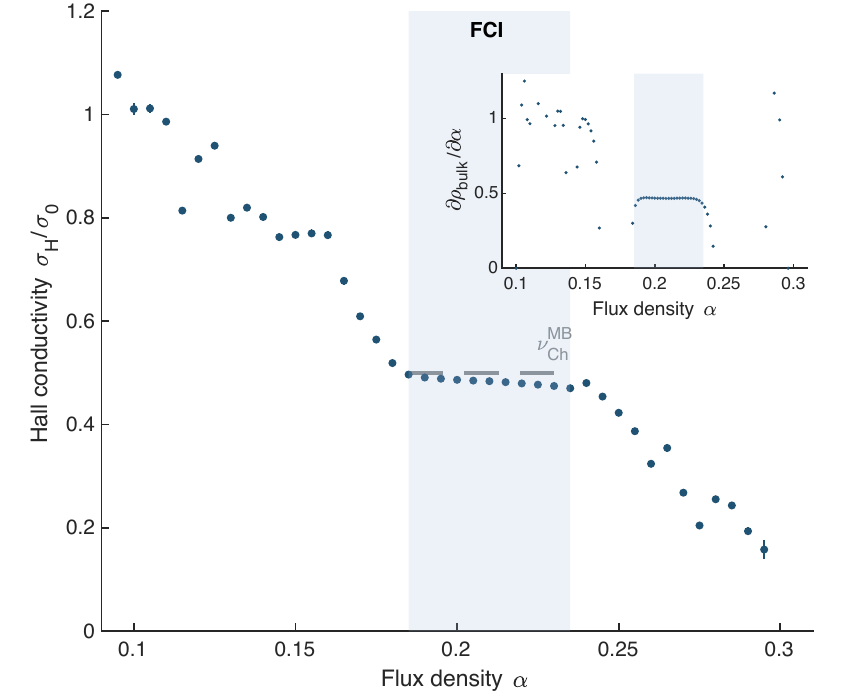}
\caption{Hall drift from $N\!=\!10$ hardcore bosons initially contained in a circular box of $N_s=120$ sites in the Harper-Hofstadter-Hubbard model, obtained using DMRG. The Hall conductivity was extracted from the COM Hall drift upon releasing the ground state into a square with $256$ sites and applying a force $F\!=\!0.04J/d$. The bulk density $\rho_{\text{bulk}}$ was evaluated as the average density in a circle comprising $16$ sites; fit error bars reflect the 95$\%$ confidence interval when extracting the Hall velocity $v_{\perp}$. The inset is a reminder of the Streda analysis [Section~\ref{section:streda}]. The gray dashed line indicates the quantized value expected in the thermodynamic limit, $\sigma_\text{H}/\sigma_0\!=\!1/2$, as dictated by the many-body Chern number.
}
\label{figsuppDMRG}
\end{figure}

We used DMRG to extend our results to larger system sizes (with $N\!=\!5$ to $N\!=\!10$ bosons); methodological details of the DMRG simulations are given in Appendix B. We simulated the Hall drift of up to $N\!=\!10$ hardcore bosons, initially prepared in a circular box of $N_s\!=\!120$ sites and released into a square box of $256$ sites; see Fig.~\ref{figsuppDMRG}. The particle density in the ground state [Fig.~\ref{figStreda}(a)] depicts a plateau in the central region; as a result, the value of $\rho_{\text{bulk}}$ is relatively insensitive to the radius $r_{\text{bulk}}$ chosen for its evaluation. In order to extract $\sigma_{\text{H}}$ from the COM Hall drift [Eq.~\eqref{eq:extract}], we evaluated $\rho_{\text{bulk}}$ as the average density in a circle of $16$ sites; the results are shown in Fig.~\ref{figsuppDMRG}. As expected for a $\nu\!=\!1/2$ FCI phase, the value of the Hall conductivity on the plateau approaches the many-body Chern number $\sigma_{\text{H}}/\sigma_0\!=\! 0.5$. We point out that the approached quantization is already accurate up to $\simeq 5\%$ for this system of $N\!=\!10$ bosons.\\

\subsection{Minimal configurations: 3 and 4 bosons}

We sought for minimal configurations that exhibit Hall-plateau signatures of the FCI state, by simulating the Hall drift protocol for various system sizes $N_s$ and number of bosons $N$. For both $N\!=\!3$ and $N\!=\!4$, we hereby show the exact-diagonalization results obtained for the smallest system size where Hall-drift signatures of a $\nu\!=\!1/2$ FCI state were found:~$N_s\!=\!40$ for $N\!=\!3$ and $N_s\!=\!49$ for $N\!=\!4$. 

Figure~\ref{fig1supp} shows the ground-state properties and the center-of-mass Hall drift for $N\!=\!4$ bosons and $N_s\!=\!49$ sites.
Based on the low-energy spectra and entanglement spectroscopy, we find clear signatures of the $\nu\!=\!1/2$ FCI state for $0.18\!\leq\!\alpha\!\leq\!0.29$. We point out that this shift of the FCI phase towards larger values of $\alpha$ compared to the case presented in the main text ($N\!=\!4$, $N_s\!=\!60$, where the FCI regime corresponds to $0.15\!\leq\!\alpha\!\leq\!0.25$; see Fig.~2 in the main text) is consistent with a larger particle density $\rho\!=\!N/N_s$. 

We observe an emergent Hall plateau for $\alpha \geq 0.18$. We note that its width is slightly smaller than the FCI region; as shown in the main text (using $N_s\!=\!60$ sites for $N\!=\!4$ bosons), this discrepancy is reduced by increasing the system size, which suggests that it is due to the smallness of this minimal setting ($N_s\!=\!49$ sites for $N\!=\!4$ bosons). We note that additional plateau features appear for each gap opening in the finite-size energy spectrum.

For $N\!=\!3$ bosons, the particle entanglement spectrum (PES) cannot provide a topological signature of the FCI. Indeed the PES that results from the only available particle bipartition ($2\!+\!1$) corresponds to the spectrum of the single-particle density matrix (which cannot probe topological order). Nevertheless, for $N\!=\!3$ bosons and $N_s\!=\!40$ sites, the low-energy many-body spectrum and the orbital occupation are compatible with a FCI state in the flux window $0.16\!\leq\!\alpha\!\leq\!0.28$, where the Hall drift simulation reveals an emergent Hall plateau; see Fig.~\ref{fig2supp}.

\section{Finite-size effects in non-interacting fermions}\label{app:local}

In this Appendix, we discuss the deviation between the Hall plateau extracted from the drift of non-interacting fermions $\sigma_\text{H}$ and the quantized value $\sigma_\text{H}/\sigma_0\!=\!\nu_{\text{Ch}}^{\text{MB}}\!\in\!\mathbb{Z}$ expected in the thermodynamic limit, where $\nu_{\text{Ch}}^{\text{MB}}$ is the many-body Chern number of the prepared insulating state; see Fig.~\ref{fig3}.

\begin{figure}[h!]
\includegraphics[width = \linewidth]{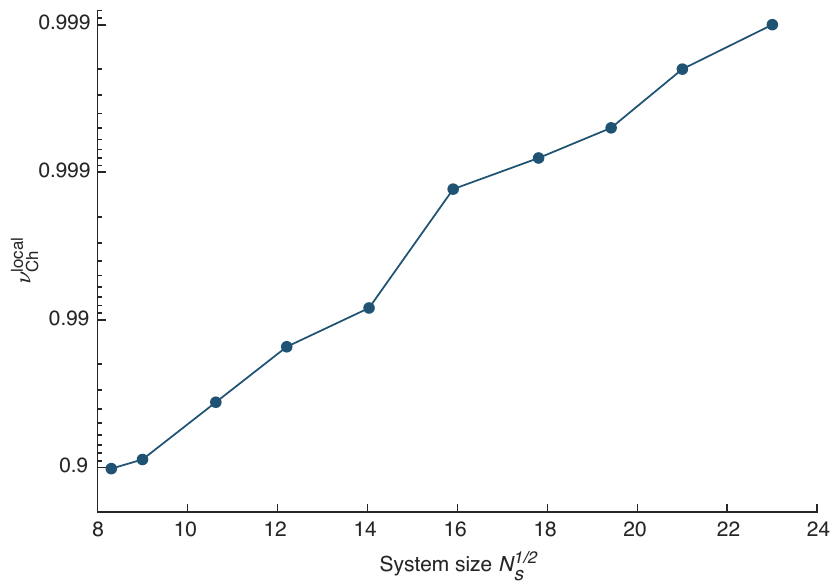}
\caption{Local Chern number as a function of the system size ($\sqrt{N_s}$, where $N_s$ is the number of lattice sites), for non-interacting fermions in the Harper-Hofstadter model in a box, at flux density $\alpha\!=\!1/4$ and total particle density $\rho\!\approx\!1/4$, i.e.~a completely filled lowest energy band. The local Chern number is obtained by averaging the local Chern marker over $29$ central sites. The value $\nu_{\text{Ch}}^{\text{local}}\!\approx\!0.9$ for $N_s\!=\!81$ lattice sites corresponds to the system configuration in Fig.~\ref{fig3}.
}
\label{LocalChern}
\end{figure}

To confirm the finite-size origin of this deviation, we have evaluated the local real-space Chern number~\cite{bianco2011mapping,tran2015topological} for the same number of fermions ($N\!=\!20$), in the same initial box of $N_s\!=\!81$ sites, and a fixed flux density $\alpha\!=\!0.25$. By averaging the local Chern marker over $29$ sites, located in a circular region at the center of the bulk, we have obtained that the local real-space Chern number matches the value of the extracted Hall conductivity, i.e.~$\nu_{\text{Ch}}^{\text{local}}\!\approx\!\sigma_\text{H}/\sigma_0\!\approx\!0.9$. We have further calculated this local real-space Chern number for increasing system sizes while fixing the particle density $\rho\!\approx\!1/4$ and flux $\alpha\!=\!0.25$ (i.e.~keeping the lowest energy band completely filled). As shown in Fig.~\ref{LocalChern}, the local topological index converges towards the quantized value $\nu_{\text{Ch}}^{\text{local}}\!\rightarrow\!1$. This behavior is reminiscent of the convergence of the non-integrated many-body Chern number under periodic boundary conditions~\cite{Kudo}; we note that these two quantities should indeed become equivalent in the thermodynamic limit.

\section{Single-particle Hofstadter spectrum in a small circular box}\label{app:spectrum}

\begin{figure}[h!]
\includegraphics[width = \linewidth]{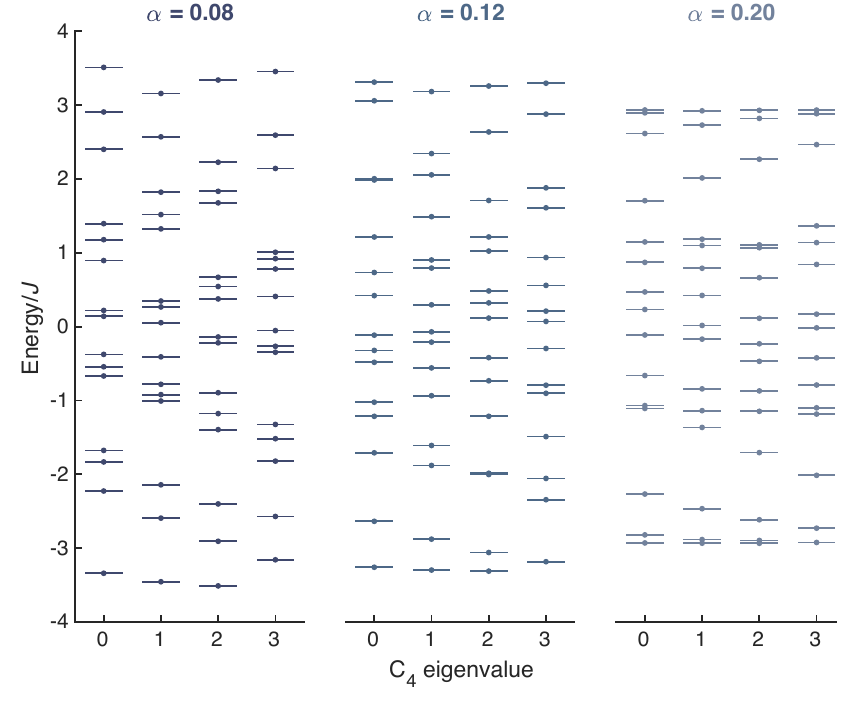}
\caption{Single-particle spectrum for $N_s=60$ sites and flux density $\alpha = 0.08$, $0.12$ and $0.2$, labeled according to the eigenvalues of the $C_4$-rotation operator. The occupation of each of these orbitals in the $N\!=\!4$ hardcore-boson configuration is given in Fig.~2(c) of the main text.
}
\label{fig3supp}
\end{figure}

\begin{figure*}[h!]
\includegraphics[width = \linewidth]{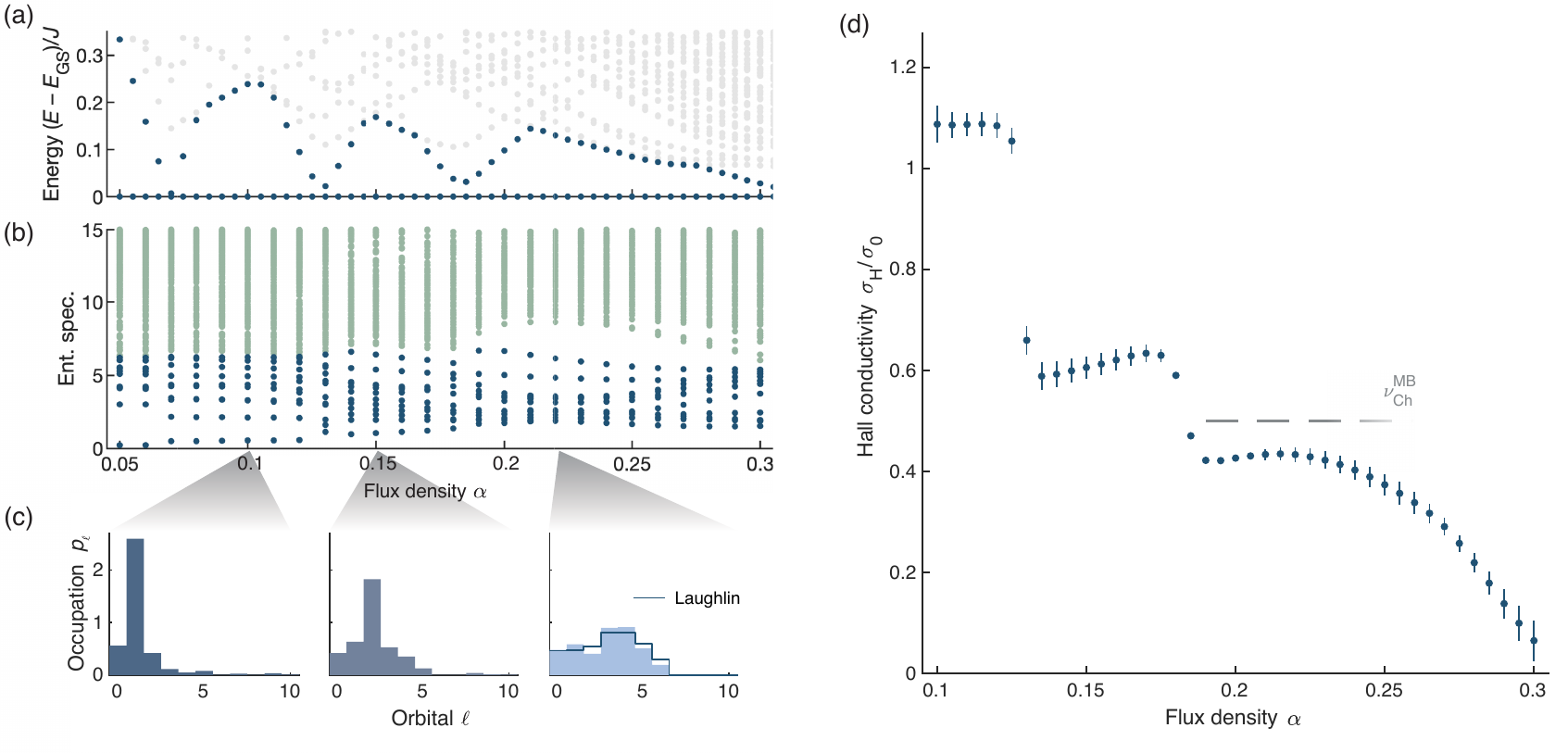}
\caption{Properties of a system of $N\!=\!4$ hardcore bosons in a circular box of $N_s=49$ sites in the Harper-Hofstadter-Hubbard model. The left column shows the characterization of the ground state through static signatures (following Fig.~2 of the main text): (a) Energy spectrum; (b) Particle Entanglement Spectrum; (c) Occupation of the single-particle orbitals in the ground state, in increasing energy order (the line shows the orbital occupation for the $N\!=\!4$ Laughlin state on the disk, where the orbitals are sorted in increasing angular momentum); (d) Hall conductivity as extracted from the COM Hall drift upon releasing the ground state into a circle with $113$ sites and applying a force $F\!=\!0.0001J/d$. The bulk density $\rho_{\text{bulk}}$ was evaluated as the average particle density within a circle comprising $9$ sites ($r_{\text{bulk}} = 0.5r_0$). The gray dashed line indicates the quantized value expected in the thermodynamic limit, $\sigma_\text{H}/\sigma_0\!=\!1/2$, as dictated by the many-body Chern number.
}
\label{fig1supp}
\end{figure*}

\begin{figure*}[h!]
\includegraphics[width = \linewidth]{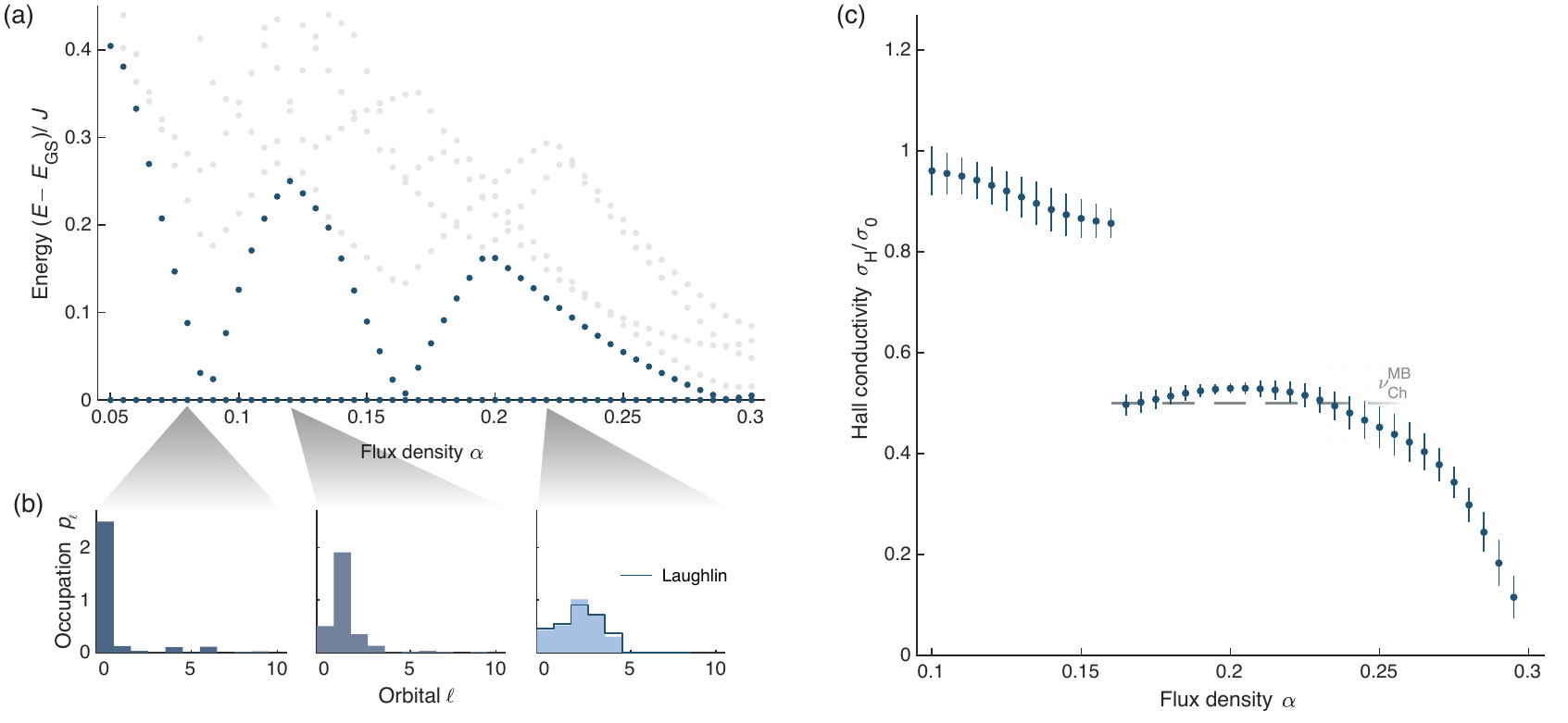}
\caption{Properties of a system of $N\!=\!3$ hardcore bosons in an elliptic box of $N_s=40$ sites in the Harper-Hofstadter-Hubbard model. The left column shows the characterization of the ground state through static signatures: (a) Energy spectrum; (b) Occupation of the single-particle orbitals in the ground state, in increasing energy order (the line shows the orbital occupation for the $N=3$ Laughlin state on the disk, where the orbitals are sorted in increasing angular momentum); (c) Hall conductivity as extracted from the COM Hall drift upon releasing the ground state into an ellipse with $100$ sites and applying a force $F\!=\!0.0001J/d$. The bulk density $\rho_{\text{bulk}}$ was evaluated as the average particle density within a circle comprising $8$ sites ($r_{\text{bulk}} = 0.5r_0$). The gray dashed line indicates the quantized value expected in the thermodynamic limit, $\sigma_\text{H}/\sigma_0\!=\!1/2$, as dictated by the many-body Chern number.
}
\label{fig2supp}
\end{figure*}

We show the single-particle Hofstadter spectrum for a lattice of $N_s\!=\!60$ sites in Fig.~\ref{fig3supp}, for three representative values of the flux $\alpha$. To label the eigenstates, we use the eigenvalues of the $C_4$-rotation operator; in this system with discrete rotational symmetry~\cite{powell2011bogoliubov,ozawa2015momentum}, they are equivalent to the angular momentum modulo $4$. 

Considering this lattice of $N_s\!=\!60$ sites and a flux $\alpha\!\approx\! 0.2$, we have shown that the ground state of $N\!=\!4$ hardcore bosons can be identified as a $\nu\!=\!1/2$ FCI ground state (see main text). In this setting, only the $7$ lowest-energy orbitals have a substantial occupation in this ground state [see Fig.2(c) of the main text]. In Fig.~\ref{fig3supp}, we show that these $7$ orbitals form the nearly-flat lowest band of the single-particle spectrum at $\alpha\!=\!0.2$.

\section{Methods}\label{methods}

\subsection{Ramps used in numerical calculations}
In the Hall-drift protocol described in the main text, the force $F(t)$ and the tunneling terms $J(r, t)$ are slowly ramped up until reaching a stationary regime at $t = \tau_{\mathrm{ramp}}$; here $r$ denotes the radial coordinate on the 2D plane. In our numerical calculations, the force is ramped up according to the first-order smoothstep function
\begin{align}
F(t) = & F \left(3 \left(\frac{t}{\tau_{\mathrm{ramp}}}\right)^2  - 2\left(\frac{t}{\tau_{\mathrm{ramp}}}\right)^3\right) ,\label{eq: ramp F}
\end{align}
while we have used an exponential ramp for the tunneling terms, 
\begin{align}
J(r, t) = & J \exp\left(-\frac{r - r_0}{r_1}  \left(\frac{\tau_{\mathrm{ramp}}}{t} - 1\right)\right) ,\label{eq: ramp J}
\end{align}
where $r_0$ is the radius of the small circular box where the initial state is prepared, and $r_1$ is the radius of the larger circular box (i.e.~the simulation box) into which it is released. These ramps were used both for the hardcore-boson (exact diagonalization and DMRG) and non-interacting-fermion cases. We expect that the exact form of the ramps should not be crucial in view of reaching a stationary regime, as long as they are smooth enough.
Besides, we have verified that the linear-response regime is reached when using a weak force $F\!\lesssim\!0.04J/d$ in the hardcore bosons simulations. 


\subsection{Particle entanglement spectrum}
We obtain the ground state $\ket{\Psi_{\mathrm{GS}}}$ of the HHH model using exact diagonalization. Here, we consider $N\!=\!4$ hardcore bosons in a box of $N_s\!=\!60$ sites. We consider a bipartition of the system with $N_A\! =\! 2$ (resp. $N_B\!=\! N - N_A \!=\!2$) bosons in $N_s$ sites in subsystem $A$ (resp. $B$). The particle entanglement spectrum~\cite{sterdyniak-PhysRevLett.106.100405} (PES) is the spectrum of $-\log\hat{\rho}_A$, where $\hat{\rho}_A\!=\! \mathrm{Tr}_B\hat{\rho}$ is the reduced density matrix obtained by tracing $\hat{\rho} \!=\!\ket{\Psi_{\mathrm{GS}}}\!\bra{\Psi_{\mathrm{GS}}}$ over the subsystem $B$. \\


\subsection{Exact diagonalization: ground state and time-evolution calculations}
For systems with $N\leq 4$ hardcore bosons, our numerical results were obtained using exact diagonalization.
To obtain the initial state, we first calculate the ground state $\ket{\Psi_{\mathrm{GS}}}$ of the HHH model in a small circular box of $N_s$ sites. This state is then embedded in a larger circular box, by performing the direct product of $\ket{\Psi_{\mathrm{GS}}}$ with several empty sites. 
The state $\ket{\Psi(t+\delta_t)}$ at time $t+\delta_t$ is obtained by applying the unitary time-evolution operator $\hat{U}(t)\!=\!e^{-i\hat{H}(t)\delta_t}$ onto $\ket{\Psi(t)}$ for a small time interval $\delta_t\!=\! 0.05J^{-1}$; here $\hat{H}(t)$ is the Hamiltonian acting on $N$ bosons in the large circular box; it corresponds to the HHH Hamiltonian in Eq.~\eqref{eq: time independent H} with uniform tunneling amplitude $J$ in the inner circular region, and tunneling amplitude $J(r, t)$ in the outer region; it also contains the  potential gradient realizing the uniform force $F(t)$. During the ramp, $0\!\leq\!t\!<\!\tau_{\text{ramp}}\!=\!8 J^{-1}$, the time-dependent quantities $F(t)$ and $J(r, t)$ are given by Eqs~\eqref{eq: ramp F}-\eqref{eq: ramp J}. After the ramp, $F(t)\!=\!F$ is constant and the tunneling amplitude is constant and uniform throughout the entire system. The time-evolution operator $\hat{U}(t)$ is evaluated using a series expansion with accuracy $10^{-15}$.\\

\subsection{DMRG: ground state and time-evolution calculations}
For systems with more than $N\!=\!4$ particles, where exact diagonalization becomes prohibitively costly, we have used DMRG. We first calculate the ground state $\ket{\Psi_{\mathrm{GS}}}$ of the HHH model in a small circular box of $N_s$. This box is embedded from the start in a larger rectangular box, but the bosons are initially confined to the small box due to the absence of tunneling terms to the outer sites. The size of the rectangular box is chosen to be large enough to permit a stationary regime during the Hall drift protocol. For the Hall drift protocol, we performed the time-evolution using the matrix product operator algorithm introduced in Ref.~\onlinecite{zaletel2015time} with the WII expression for the time-evolution operator (see  Eq.  10  therein), and a time step $\delta_t \!=\! 0.02J^{-1}$. We verified that the error due to this finite time step was smaller than the linear fit error associated with the extraction of the Hall velocity $v_{\perp}$.
For $N\!=\!10$ and $N_s\!=\!120$, the results shown here were obtained with a bond dimension $\chi\!=\!500$; we have verified that using $\chi\!=\!800$ only resulted in a negligible difference of the COM motion.
\\

\subsection{The case of non-interacting fermions}
We first calculate the ground state of $N$ fermions in the non-interacting Harper-Hofstadter model, within a circular box containing $N_s$ sites. This sets our initial state $\vert \Psi_{\text{GS}} \rangle$, for a given flux density $\alpha$. In our calculations, we set $N\!\approx\!N_s/4$ so that the particle density is close to quarter filling $\rho\!\approx\!1/4$. We then obtain the time evolution operator $\hat U_{\text{ramp}}$ describing the ramp, during which both the force and the tunneling into the larger lattice are activated; since the corresponding Hamiltonian is explicitly time-dependent, we discretize the time evolution in small time steps $\Delta_t\!=\!0.1J^{-1}$; the full duration of the ramp is $\tau_{\text{ramp}}\!=\!15J^{-1}$. From this, we calculate the state at the end of the ramp through $\vert \Psi_{\text{ramp}} \rangle\!=\!\hat U_{\text{ramp}}\vert \Psi_{\text{GS}} \rangle$. Finally, the center-of-mass of the system is monitored by calculating the time-evolution of the state after the ramp, $\vert \Psi (t) \rangle\!=\!\hat U_{\text{hold}}(t)\vert \Psi_{\text{ramp}} \rangle$, in the presence of the constant force.

\end{document}